\documentclass[10pt,aps,prb,twocolumn,amsmath,amssymb,superscriptaddress,longbibliography]{revtex4-1}
\usepackage[pdftex, colorlinks=true, linkcolor=blue, citecolor=blue, urlcolor=blue]{hyperref}
\usepackage{graphicx}
\usepackage{aurical}
\usepackage[T1]{fontenc}
\usepackage{units}
\usepackage{siunitx}
\usepackage[version=3]{mhchem}
\usepackage{xcolor}
\usepackage{bm}
\definecolor{darkgreen}{rgb}{0.0, 0.5, 0.0}
\usepackage{mathtools}

\DeclarePairedDelimiterX\braket[2]{\langle}{\rangle}{#1 \delimsize\vert #2}

\graphicspath{{./figures/}}
\DeclareGraphicsExtensions{.png,.pdf}

\newcommand{\kaust}{Physical Science and Engineering Division (PSE), King Abdullah University of Science and Technology (KAUST), Thuwal 23955-6900, Saudi Arabia}
\newcommand{\mrs}{Aix-Marseille Univ, CNRS, CINaM, Marseille, France}

\begin{document}
	
	\title{Spin-orbit coupling induced ultra-high harmonic generation from magnetic dynamics}
	\author{Ousmane Ly}
	\email[]{ousmane.ly@kaust.edu.sa}
	\affiliation{\kaust}
	\author{Aurelien Manchon}
	\email[]{aurelien.manchon@cinam.univ-mrs.fr}
	\affiliation{\kaust}
	\affiliation{\mrs}

	\maketitle
	
	{\bf The recent boost in data transfer rates puts a daring strain on information technology. Sustaining such a growth rate requires the development of sources, detectors and systems working  in the so-called TeraHertz (THz) gap covering the frequency window from 0.1 to 10 THz (1 THz = 10$^{12}$~Hz). This gap represents a challenge for conventional electronic devices due to carrier transit delays ($\sim$1-10ps), as well as for photonic devices due to thermal fluctuations (300K$\sim$6THz). Nonetheless, designing efficient, room-temperature THz sources would constitute a key enabler to applications spanning from high-resolution imaging to extreme wide band wireless communication. Whereas high-harmonic generation in solid is usually limited to less than ten harmonics, broadband THz emission  has been demonstrated using laser-induced superdiffusive spin currents in magnetic bilayers composed of a ferromagnet deposited on top of a noble metal.  While promising, this technique presents the major disadvantage of necessitating optical pumping and hence lacks scalability. Here, we demonstrate that extremely high harmonic emission can be achieved by exploiting conventional spin pumping, without the need of optical excitation. 
		We show that when the spin-orbit coupling strength  is close to  the s-d exchange energy, a strongly non linear regime resulting from resonant spin flip scattering occurs leading to the generation of a thousand of harmonics at realistic antiferromagnetic precession frequencies, thereby enhancing both spin and charge dynamics by two orders of magnitude, and allowing for an emission at frequencies above 300 THz.}
	
	The grand challenge posed by the THz gap is that it is located on the high-end of electronic processes and on the low-end of optical excitations. It is therefore difficult to emit THz electromagnetic field using purely electronic mechanisms because the scattering time of the electronic carriers is typically in the THz range. A successful strategy exploited in quantum cascade lasers is to use semiconductor superlattices with small gaps in order to generate the required frequency. Since the efficiency is low, one needs to multiply the number of gaps (hence, the quantum cascade) and work at low temperature to quench thermal fluctuations. Another strategy is to exploit optical rectification of a femtosecond laser pulse in semiconductors like ZnTe. An alternative approach is to exploit an optically-generated ultrafast spin current, i.e. a charge-neutral current carrying spin angular momentum. In this geometry, a femtosecond laser pulse impinges on the surface of a thin ferromagnetic film and excites a superdiffusive spin current. This spin current then penetrates into an adjacent layer possessing large spin-orbit coupling and is subsequently converted into a charge current via either spin Hall effect \cite{Sinova2015} or Rashba-Edelstein effect \cite{Edelstein1990}, depending on the  considered heterostructure. This apparatus enables the conversion of a femtosecond laser pulse into a THz electromagnetic field \cite{Kampfrath2013,Huisman2016}. Although promising this solution relies on optical pumping and therefore lacks scalability. 
	
	In this work, we demonstrate that extremely efficient and ultrafast THz emission can be obtained by exploiting the internal dynamics of the spin-to-charge conversion to generate high harmonics in a device solely powered by conventional magnetic resonance. The driving force is given by spin pumping \cite{Brataas2002,Saitoh2006}, a widely used technique to generate pure spin currents out of ferromagnetic resonance [see Fig. \ref{fig:sketch}(a)]. 
	
\begin{figure}
		\includegraphics[width=8cm]{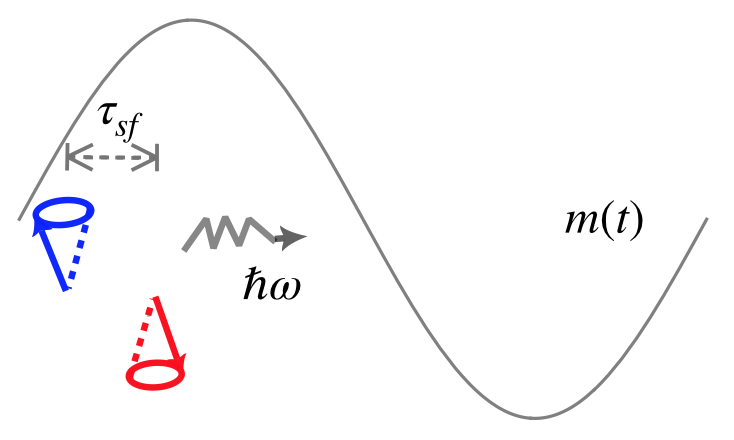}
		\caption{\textbf{Spin-flip scattering driven high harmonic generation:} In the presence of a precessing magnetization order (gray line)  the spin orbit interaction leads to spin-flip scattering.  Subsequently,  a phase corresponding to the fundamental frequency is accumulated in the electronic wave-function. At adiabatic magnetization dynamics, the spin flip time ($\tau_{sf}$) becomes very small compared to the period of the precessing magnetic order, therefore higher harmonics appear as a consequence of subsequent spin flip scattering events during one fundamental cycle.}
		\label{fig1}
	\end{figure}

\begin{figure}
		\includegraphics[width=8cm]{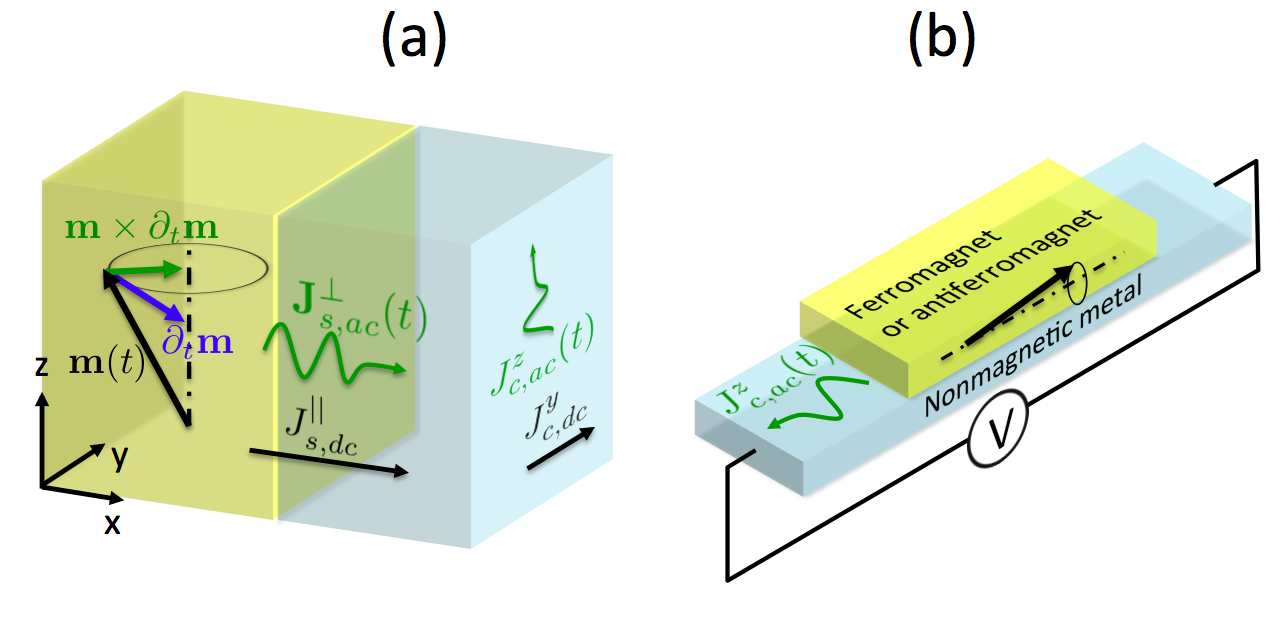}
		\caption{\textbf{Illustration of the spin pumping setup:} \textbf{a} Schematics of the d.c. and a.c. spin pumping: a magnetic order precessing around its rest direction ${\bf z}$ induces a spin current composed of a d.c. component, $J_{s,dc}^{||}$ (black), and an a.c. (green) component, $J_{s,ac}^{\bot}$, whose spin polarization is aligned on ${\bf m}\times\partial_t{\bf m}$. We neglect the contribution from the imaginary part of the spin mixing conductance ($\sim\partial_t{\bf m}$). Via spin-orbit coupling, these spin currents are converted into a d.c. and an a.c. charge current flowing along ${\bf y}$ and ${\bf z}$, respectively. \textbf{b} Corresponding simulation setup: the pumped charge current is collected along the direction about which the magnetic order precesses.}
		\label{fig:sketch}
	\end{figure}

	The key ingredient of this phenomenon relies on the spin scattering events induced by spin-orbit coupling in the presence of a precessing magnetic order. This leads to strongly non-linear spin and charge responses. 
	A phenomenological explanation of the effect  can be stated as follows: 
	while the magnetic order is precessing with a given initial frequency the electronic spin undergoes spin-flip scattering events in the presence of spin-orbit interaction (see Fig. \ref{fig1}). Subsequently, a phase corresponding to the dynamic frequency is accumulated as a consequence of angular momentum conservation. 
	This is in close analogy with the high harmonic generation mechanism from gaseous media under Laser excitation \cite{Corkum1993, Corkum2007, Midorikawa2011}, where  the electron wave-packets undergo harmonic emission  as they re-collide to their parent atoms after a fraction of the Laser oscillation cycle. When the magnetization dynamics operates at small frequencies the emission of a huge number of harmonics is predicted. 
	Interestingly, our numerical simulations suggest the existence of a resonance condition associated to maximally excited high-harmonics. This corresponds to the regime where the s-d exchange energy is very close to the spin-orbit splitting. 
	While the effect is not restricted to a particular type of spin-orbit interaction, the Rashba-like spin-orbit coupling constitutes a central paradigm to demonstrate the proposed effect.  In fact, Rashba spin-orbit interaction has been found in a broad range of magnetic interfaces, from transition metal interfaces \cite{Saitoh2006}, to the surface of topological insulators \cite{Shiomi2014} or in oxide heterostructures \cite{Caviglia2010}, and has the major advantage of being electrically tunable \cite{Vaz2019}, offering a powerful means to control the high harmonic generation.
	
	Spin pumping \cite{Brataas2002} is an adiabatic process by which a precessing magnetic order ${\bf m}(t)$ (ferromagnetic or antiferromagnetic alike) injects a spin current into an adjacent metallic layer. This spin current possesses two components, a rectified one whose spin polarization is aligned along the precession axis ${\cal J}_s\sim \langle {\bf m}\rangle$, and an oscillating contribution whose spin polarization is aligned perpendicular to the precession axis ${\cal J}_s\sim{\bf m}\times\partial_t{\bf m}$. In a conventional spin pumping experiment, once injected in the normal metal, this spin current is converted into a charge current via spin-orbit coupling \cite{Saitoh2006, Wei2014, Rojas-Sanchez2013b}. This spin-to-charge conversion does not affect the spin dynamics itself as long as the spin-orbit coupling is negligible compared to the s-d exchange, and thereby only generates a harmonic charge current. In the presence of Rashba spin-orbit coupling though, successive spin flip events accompanied by harmonic emission occur leading to the appearance of higher frequencies in the charge current signal.  When the spin-orbit coupling energy becomes comparable to the s-d exchange parameter, the full harmonic spectrum is excited with extremely strong amplitudes.

	To demonstrate this effect, we consider the system depicted on Fig. \ref{fig:sketch}(b), where a magnetic structure is attached to a two-dimensional metal with Rashba spin-orbit coupling. This system is modeled by a tight-binding Hamiltonian and we numerically solve the time-dependent Schr\"odinger equation and determine the different time-dependent propagating wave functions $\Psi_{\eta,\varepsilon}(t)$ coming from each lead $\eta$ at energy $\varepsilon$ using the state-of-the-art time-dependent quantum transport package developed in Refs. \onlinecite{Weston2016, Gaury2014, Kloss2021} (see Methods). It is important to note that the present effect is independent on the nature of the magnetic resonance and is only governed by the ratio of the spin-orbit coupling strength to the dynamic frequency. 
	In fact, the high harmonic generation is obtained for both ferromagnetic and antiferromagnetic resonances (see Ref. \onlinecite{SuppMat}). Nonetheless, in the numerical method we use, the simulation time is set by the electron's energy and is of the order of the hopping parameter $\gamma$ (typically of the order of 0.1 eV). In order to keep the computational cost reasonable, we have to consider a magnetic system whose resonance frequency is only two orders of magnitude smaller than the internal dynamics of the conduction spin. To comply with the numerical constraints, we therefore consider antiferromagnetic resonance (typically $\sim$meV) rather than ferromagnetic resonance (typically $\sim\mu$eV). We stress out that this does not affect the generality of our results.
	
	In antiferromagnetic resonance the two sublattice magnetization vectors, $\mathbf{m}_1$ and $\mathbf{m}_2$, undergo different precession modes of opposite chirality \cite{Cheng2014c}. Without loss of generality, we consider the right-handed polarization of the order parameter, in which case both sublattice vectors rotate anticlockwise \cite{Keffer1952}. As a result, the antiferromagnet order parameter $\mathbf{n}=(\mathbf{m}_1-\mathbf{m}_2)/2$ precesses with the same chirality and a non-zero, albeit small, in-plane magnetization $\mathbf{m}=(\mathbf{m}_1+\mathbf{m}_2)/2$ develops. As the sublattice magnetizations precess in time, the spin current pumped out of the antiferromagnet reads\cite{Cheng2014c}, 
	\begin{equation}
	\label{eq:totalJs}
	J_s = \Re\{g^{\uparrow\downarrow}\} \left( {\mathbf{m}}\times \partial_t\mathbf{m} +\mathbf{n}\times \partial_t \mathbf{n}\right )-\Im\{g^{\uparrow\downarrow}\} \partial_t\mathbf{m},
	\end{equation}
	where $g^{\uparrow\downarrow}$ is the interfacial spin mixing conductance. The symbols $ \Re\{...\} $ and $\Im\{...\} $ stand for real and imaginary parts, respectively. In contrast to the ferromagnetic case, the spin current pumped out of an antiferromagnetic metal decomposes into a couple of a.c. spin currents: a ferromagnetic like contribution $J_m^s$ along the precession cone axis, given by the first and last terms of equation \eqref{eq:totalJs}, and a staggered contribution $J_n^s$ rotating in the xy plane, given by the second term of equation \eqref{eq:totalJs}. In the region subjected to the spin-orbit interaction, the two a.c. spin current contributions are converted into a.c. charge currents in both y and z directions according to the inverse spin Hall effect \cite{Wei2014}. 
	
	\begin{figure}
		\includegraphics[width=8cm]{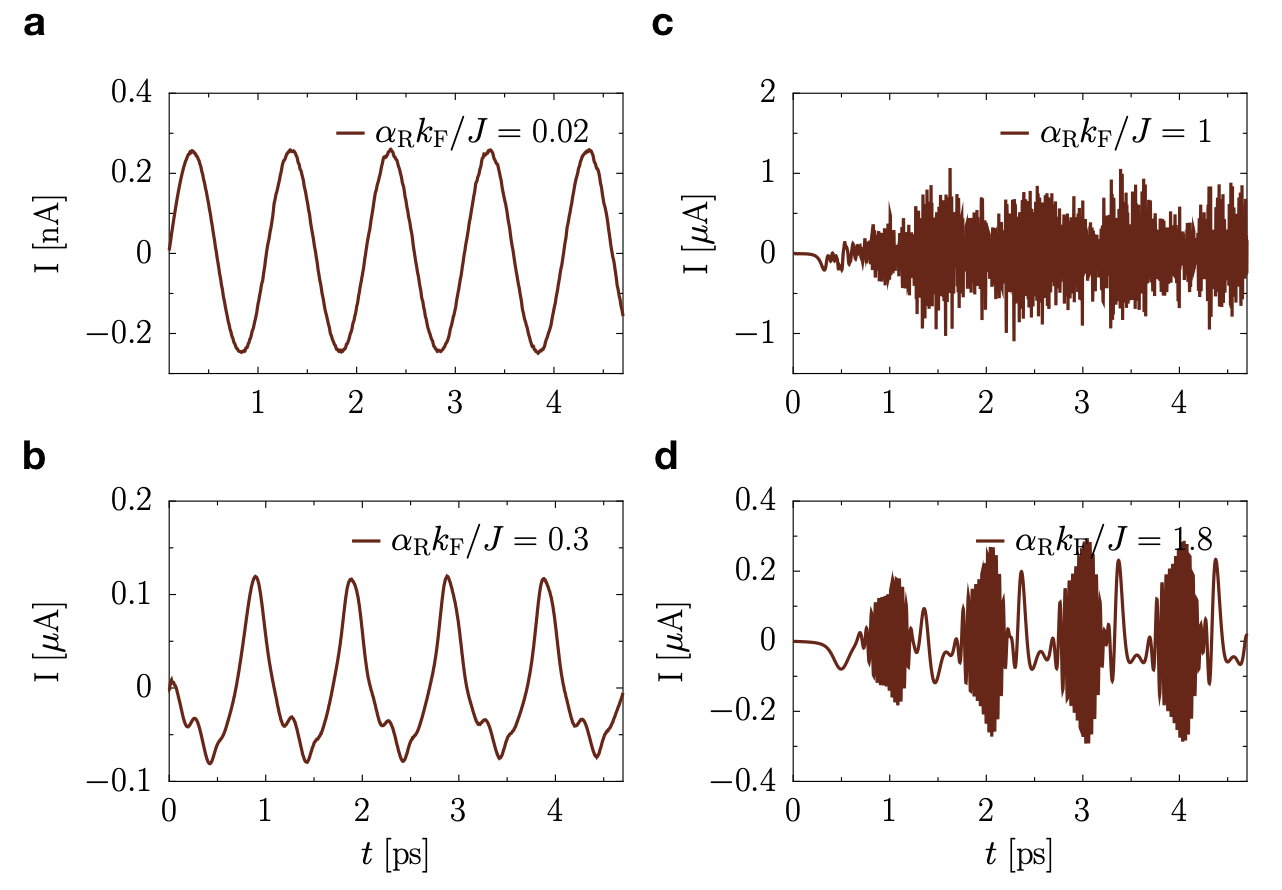}
		\caption{\textbf{Time domain current signal vs Rashba strength:} Time dependent currents for different spin orbit coupling strengths are displayed. 
			Here, an exchange coupling of $J=500$meV is taken and the precession angle is set at  $\theta =10^{\circ}$. Considering a lattice spacing of 5 {\AA} the value of $\alpha_{\rm R}$ at $\alpha_{\rm R} k_{\rm F}/J=1$ corresponds to 3.57 eV {\AA}, which is very close to its value in Bi/Ag alloys \cite{Ast2007} or in other bismuth based topological insulators\cite{Zhang2009}.}
		\label{fig:currents}
	\end{figure}
	
	The time dependence of the collected current is displayed on Fig. \ref{fig:currents} for different Rashba strengths. We immediately identify three main regimes. For $\alpha_{\rm R} k_{\rm F}/J\ll 1$, the current response is dominated by oscillations of frequency $\omega$ [Fig. \ref{fig:currents}(a)]. When the Rashba parameter becomes comparable to the s-d exchange, $\alpha_{\rm R} k_{\rm F}/J\sim 1$, the magnitude of the a.c. current substantially increases whereas involving oscillations with higher frequencies. Upon further increasing the Rashba parameter, $\alpha_{\rm R} k_{\rm F}/J> 1$, the amplitude of the signal decreases substantially with very weak components of the higher harmonics. Further numerical simulations suggest the reappearance of the highly excited harmonics regime when $\alpha_{\rm R} k_{\rm F}/J \gg 1$. However, the underlying parameter space corresponds to unrealistic spin orbit coupling strengths. 
	
	\begin{figure}[h!]
		\includegraphics[width=8cm]{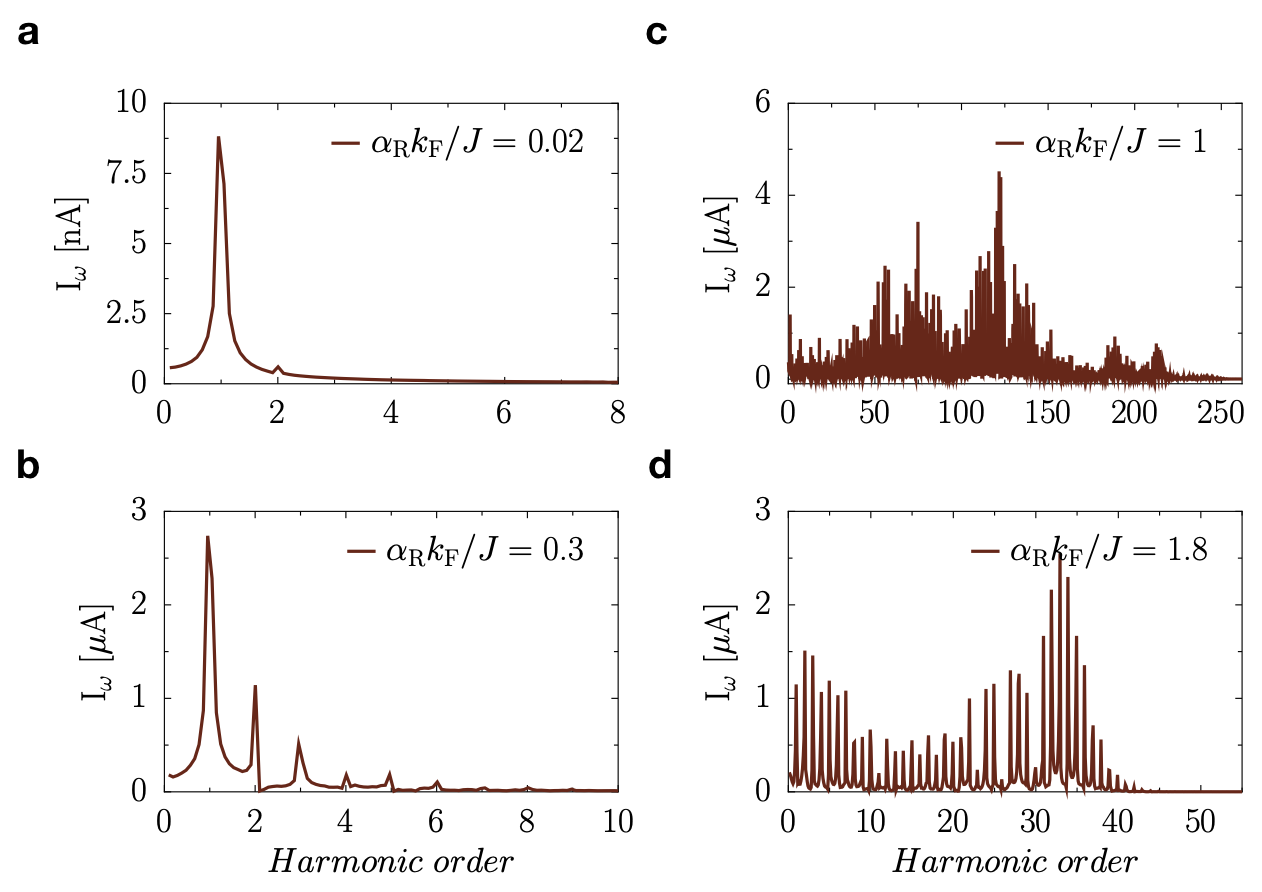}
		\caption{\textbf{Frequency domain  charge current vs Rashba strength:} Fourier amplitudes of the charge current are shown at different spin orbit coupling strengths in both perturbative and non perturbative regimes. The antiferromgantic dynamics parameters are the same as in Fig. \ref{fig:currents}.}
		\label{fig:fourier}
	\end{figure}
	
	For a more quantitative discussion, we report in Fig. \ref{fig:fourier} the Fourier transform of the charge current signals for different values of $\alpha_{\rm R}$. For $\alpha_{\rm R} k_{\rm F}/J\ll 1$ [Fig. \ref{fig:fourier}(a)], the signal exhibits only one frequency, although we do observe the appearance of the two lowest harmonics. Upon increasing the Rashba strength [Fig. \ref{fig:fourier}(b)], higher harmonics progressively emerge with decreasing amplitudes. Notice that the amplitudes of the higher harmonics are only one order of magnitude smaller than the fundamental harmonic at frequency $\omega$. For the sake of comparison, Ref. \onlinecite{Hafez2018} recently reported the optical generation of higher harmonics in graphene driven by internal electron thermalization. The authors observed harmonics up to the seventh order with amplitudes four orders of magnitude smaller than that of the fundamental mode. In contrast, our simulations demonstrate that the first few harmonics remain of the order as the fundamental frequency in this regime. The case $\alpha_{\rm R} k_{\rm F}/J\sim 1$ reported on Fig. \ref{fig:fourier}(c) is of central interest: in this regime, up to two hundred harmonics are excited, all exhibiting strong amplitude, much larger than in the single harmonic case. Using our numerical parameterization, the high harmonic spectrum extends up to more than 300 THz. Finally, when $\alpha_{\rm R} k_{\rm F}/J\ll1$ [Fig. \ref{fig:fourier}(d)], the current response is dominated by the contribution of a few lower harmonics and the overall magnitude of the signal decreases substantially. \par

	\begin{figure}
		\includegraphics[width=8cm]{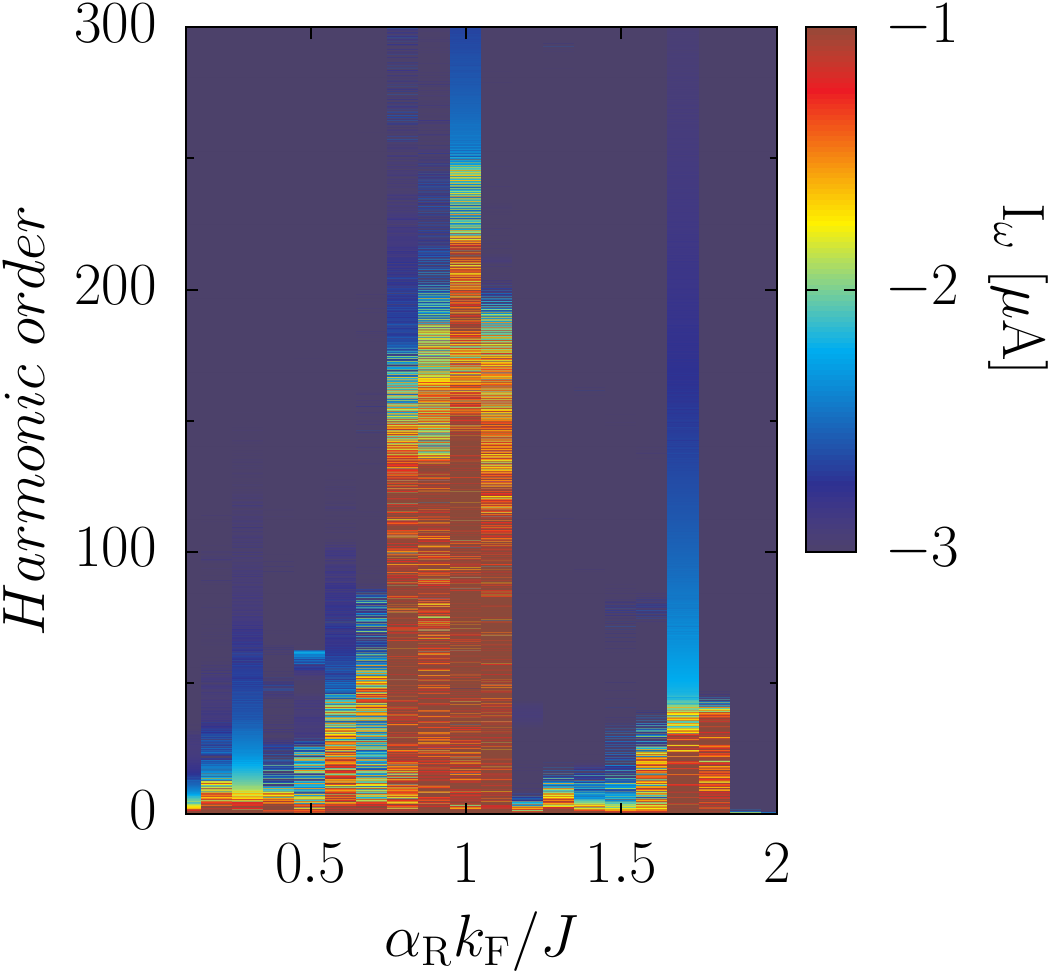}
		\caption{\textbf{Charge current amplitude of the harmonic spectrum vs spin orbit coupling strength:} The current amplitude (in logarithmic scale) is shown as a function of the Harmonic order as well as the Rashba strength. The frequency of the dynamics as well as the precession angle are the same as in Fig. \ref{fig:currents}. The figure shows four distinguishable regimes. At low $\alpha_{\rm R}$, only the few lowest harmonics appear with strong intensities. In the region around $\alpha_{\rm R} k_{\rm F}= J $  a strong response of all harmonics is observed. 
			Furthermore, when $\alpha_{\rm R} k_{\rm F}$ is bigger than $2J$ the response is also dominated by the very few lowest harmonics, with amplitudes much more smaller than in the low $\alpha_{\rm R}$ regime.} 
		\label{fig:phase}
	\end{figure}
	
\begin{figure}
		\includegraphics[width=8cm]{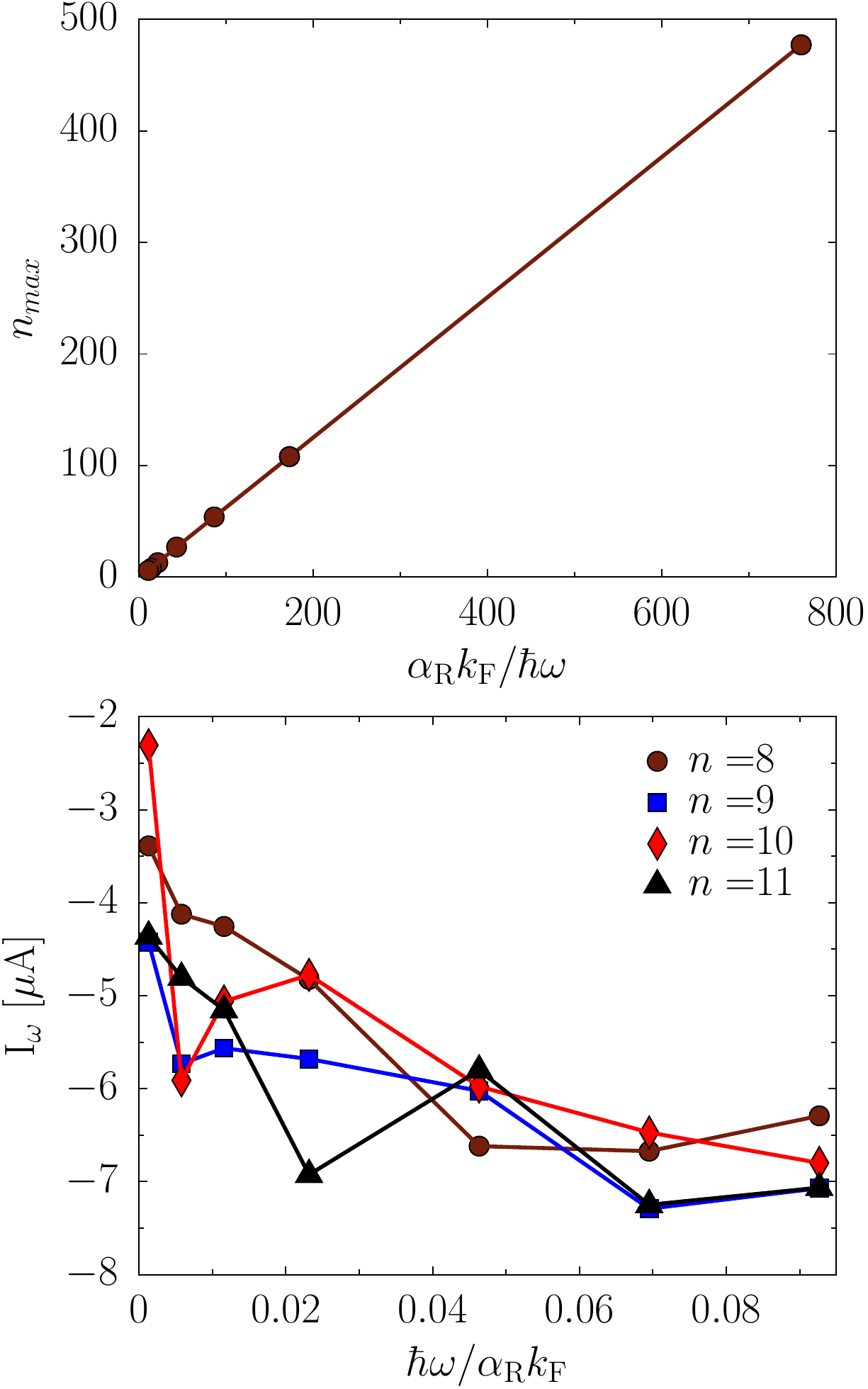}
		\caption{\textbf{Emission bandwidth and charge current amplitude vs dynamic frequency:} {\bf a} Maximum number of harmonics  as a function of the inverse of the fundamental frequency. {\bf b} Amplitude of selected harmonics vs driving frequency. The Rashba strength is tuned to the maximally excited regime.\label{fig:w}}
	\end{figure}

	In order to offer a comprehensive picture of the high harmonics generation, the current output is reported on Fig. \ref{fig:phase} as a function of the spin-orbit strength $\alpha_{\rm R} k_{\rm F}/J$ and the harmonic order, the amplitude of the harmonic being given by the color scale. The resonance regime corresponding to the ultra-high harmonic  generation is centered around $\alpha_{\rm R} k_{\rm F}/J=1$ and extends from $\alpha_{\rm R} k_{\rm F}/J\approx0.8$ to $\alpha_{\rm R} k_{\rm F}/J\approx1.2$, indicating that the effect is robust against material's parameters variations and thereby providing a wide region of tunability. 
	As a matter of fact, we observed that the signal bandwidth is proportional to the driving frequency ($\propto 1/\omega$), as shown in Fig. \ref{fig:w}(a). As demonstrated explicitly below, the amplitude of the $n$th harmonics is proportional to $\sin^n {\theta}$, $\theta$ being the cone angle of the magnetic precession. Therefore, although Fig. \ref{fig:w}(a) predicts an emission up to the 500th harmonic for the dynamic frequency considered in Fig. \ref{fig:phase}, only 300 of them exhibit strong enough amplitudes for the precession angle considered (here, $\theta=10^\circ$). At larger precession angle all the higher harmonics can be excited at resonance with amplitudes comparable to the first harmonic intensity (see Supplemental Materials\cite{SuppMat}). A similar decay of the number of the generated harmonics with respect to $\omega$ has been reported in different solid state systems under intense laser excitation \cite{Ghimire2019, Luu2015, Wu2015}. 
	
	
	To explicitly address the interplay between the magnetization dynamics parameters and spin-orbit coupling, we develop an analytical theory of the high harmonic generation with particular focus on Rashba spin-orbit coupling. 
	
	Consider the following continuous time dependent Hamiltonian,
	
	\begin{eqnarray}
	\label{eq:h}
	{\mathcal H}&=&{\mathcal H}_0(t)+{\mathcal H}_{\rm R},\\
	{\mathcal H}_0(t)&=&\frac{\hbar^2{\bf k}^2}{2\mu} +
	J \hat{{\bm\sigma}}\cdot\mathbf{m}(t)\\
	{\mathcal H}_{\rm R}&=& \alpha_{\rm R} \hat{{\bm\sigma}}\cdot({\bf z}\times{\bf k})
	\end{eqnarray} 
	where ${\mathcal H}_0$ is the single electron Hamiltonian of the magnetic system including the s-d exchange $J$,  ${\bf m}(t)$ being the time-dependent magnetic order, $\hat{{\bm\sigma}}$ is the vector of Pauli spin matrices. The constants $\mu$ and $\hbar$ are respectively, the electron mass and the reduced Planck's constant. The term ${\mathcal H}_{\rm R}$ gives the Rashba spin-orbit Hamiltonian, where $\alpha_{\rm R}$ is  the Rashba parameter  and $\bf k$ the momentum vector. 
	We consider a magnetization dynamic of the form $\mathbf{m}(t)=(\sin(\theta)\cos(\omega t), \sin(\theta)\sin(\omega t), \cos(\theta))$.
	The corresponding unperturbed spin-dependent wave functions (at $\alpha_{\rm R}=0$) read
	
	\begin{eqnarray}
	\label{eq:psi}
	\label{eq:updown}
	\Psi^{\uparrow}(t)&=&\begin{pmatrix} 
	\cos{\frac{\theta}{2}} \\
	e^{i \omega t} \sin{\frac{\theta}{2}}
	\end{pmatrix},\\
	\Psi^{\downarrow}(t)&=&\begin{pmatrix} 
	\sin{\frac{\theta}{2}} \\
	- e^{i \omega t} \cos{\frac{\theta}{2}}
	\end{pmatrix}.
	\end{eqnarray} 
	
	
	At finite  spin-orbit coupling, the exact wavefunctions evaluated at energy $\varepsilon$  can be found from the Lippmann-Schwinger equation as follows
	\begin{equation}
	\label{eq:lippmann}
	{\tilde{\Psi}^{\uparrow,\downarrow}}(t)=\Psi^{\uparrow,\downarrow}(t) + \int {\mathcal G}^{\uparrow,\downarrow}(t,t') H_R{\tilde{\Psi}^{\uparrow,\downarrow}}(t') \ dt' ,
	\end{equation}
	where $\Psi^{\uparrow,\downarrow} (t)$ and  ${\mathcal G}^{\uparrow,\downarrow}(t,t')$ are respectively, the unperturbed wave function and retarded Green function. Being interested in the behavior of the transport quantities in the time domain, we have omitted the explicit spatial dependence on both wavefunctions and Green functions. Therefore, the latter are simply given by 
	
	\begin{eqnarray}
	\label{eq:g}
	{\mathcal G}^{\uparrow}(t,t') = 
	\begin{pmatrix}
	\cos^2 {\frac{\theta}{2}} & 0 \\
	0 & e^{i \omega(t-t')}  \sin^2 {\frac{\theta}{2}}  
	\end{pmatrix} ,\\
	{\mathcal G}^{\downarrow}(t,t') = 
	\begin{pmatrix}
	\sin^2 {\frac{\theta}{2}}  & 0 \\
	0 & e^{i \omega(t-t')}   \cos^2 {\frac{\theta}{2} }    
	\end{pmatrix}.
	\end{eqnarray} 
	To obtain the $n$th order correction to the unperturbed scattering spinor, the equations \eqref{eq:g} and \eqref{eq:psi} are plugged into \eqref{eq:lippmann}. After tedious, albeit straightforward algebraic manipulations, the underlying $n$th order correction are obtained as
	
	
	
	
	
	\begin{equation}
	\label{eq:psimodd}
	\Psi^{{\uparrow,\downarrow}(n)}(t)=\pm e^{i \Theta}f_n(t) \zeta^{\uparrow,\downarrow}, 
	\end{equation}
	for odd orders, and 
	\begin{equation}
	\label{eq:psimeven}
	\Psi^{{\uparrow,\downarrow}(n)}(t)=(-1)^{\frac{n-1}{2}}e^{i\frac{\omega t}{2}} f_n(t)  \Psi^{{\uparrow,\downarrow}}(t), 
	\end{equation}
	for even $n$.
	
	The coefficients $f_n$ are given by 
	\begin{equation}
	\label{eq:fo}
	f_n(t)=\left(\frac{\alpha_{\rm R} k}{\omega}\right)^n \left(\sin {\frac{\omega t}{2}}\sin{\theta}\right)^n.
	\end{equation}
	Here, the phase $\Theta$ is defined according to $$  (k_y + ik_x)=k e^{i \Theta}$$
	and $\zeta^{\uparrow,\downarrow}$ stand  for the respective up and down components along the Bloch vector defined by the spherical angles $(\theta, \phi=-2 \Theta)$.
	
	Although the Lippmann Schwinger equation is often used to obtain perturbative responses \cite{Jalabert2010, Gorini2013}, the wavefunctions (equations (\ref{eq:psimodd}) and (\ref{eq:psimeven})) are general for any non-linear order. Therefore, in the presence of a high order cutoff as it is usually the case in high harmonic generation, these equations accurately describe the full bandwidth of the harmonic spectrum. 
	It is clear from the expressions of $\Psi^{\uparrow,\downarrow (n)}(t)$ that higher frequencies appear as $\alpha_{\rm R}$ is increased whereas the high-harmonic amplitude decreases with increasing the pumping frequency, consistently with the numerical results [Fig. \ref{fig:w} (b)]. 
	The trend of $n_{max}$ shown in Fig. \ref{fig:w}(a) suggests the possibility of generating a huge amount of higher harmonics for magnetic dynamics operating at low frequencies. 
	Considering a fundamental frequency of 0.4 THz, an ultra-high harmonic spectrum displaying more than a thousand harmonics is obtained (see Supplemental Materials \onlinecite{SuppMat}).
	
	
	In order to confirm the robustness and generality of the results reported here, we have confirmed that such a high-harmonic generation remains insensitive to the size of the system and to the number of electron modes involved in the transport \cite{SuppMat}. We have also confirmed the weak influence of electron scattering by inserting Anderson-like random impurity potential in the system. This is a very important observation because one of the bottlenecks of TeraHertz electronics is the fact that electron's momentum scattering rate typically lies in the THz gap, thereby hampering the transmission of THz currents. In our system, the ultra-high harmonic emission is governed by the electron spin, rather than its charge and therefore the underlying  spectrum is hardly influenced by electron scattering. Finally, we have also verified that the emission remains robust at room temperature \cite{SuppMat}. We observe an ultra-high harmonic emission qualitatively comparable to the zero temperature response.
	
	A crucial question we now wish to comment on is the applicability of our single-orbital model to realistic systems. As mentioned in the introduction, adiabatic spin-to-charge conversion has been demonstrated in a wide variety of heterostructures accommodating both Rashba-like spin-orbit coupling and s-d exchange. Transition metal ferromagnets interfaced with topological insulators\cite{Shiomi2014} and oxide heterostructures\cite{Vaz2019} are among the most promising structures. In these systems, Rashba-like spin-orbit coupling arises from a complex interfacial orbital hybridization scheme that is essentially overlooked in our model. Nonetheless, intense theoretical investigation on spin-orbit physics\cite{Manchon2019} (spin-orbit torque and spin-to-charge conversion) has established that in spite of its simplicity, modeling the interfacial spin-orbit coupling by an effective Rashba interaction is sufficient to properly describe the physics at stake and obtain reasonable orders of magnitude. Furthermore, the robustness of the effect at room temperature is of great relevance in terms of the applicability of the effect to real devices. \par
	
	The present effect opens appealing perspectives for high frequency emission deep into the THz gap. 
	It is worth emphasizing that the effect requires two essential ingredients: strong spin orbit coupling and oscillating magnetic order or magnetic field with inplane components. Conventional spin-to-charge conversion heterostructures are suitable platforms to harness the effect.   
	The very nature of Rashba spin-orbit coupling makes this perspective quite appealing because it is directly related to the interfacial potential drop and therefore highly sensitive to a gate voltage for instance. Whereas this electrical tuning of the Rashba strength has been demonstrated in several systems, oxide two-dimensional electron gases \cite{Lesne2016, Vaz2019, Noel2020} stand out of the most versatile system. The recent demonstration of electrical switching of Rashba coupling \cite{Noel2020} makes this perspective even more compelling. The Rashba strength can be as large as a few 100 meV, which seems reasonable as the effective s-d exchange experienced on the surface can be tuned by inserting a tunnel barrier for instance. Furthermore, a strong enhancement of spin orbit coupling strength in graphene to 80 meV has been recently reported \cite{Afzal2019}, where the spin orbit coupling strength can be controlled in a field effect transistor setup.  This provides a variety of systems in which the effect can be observed.  

	\acknowledgments
	This work was supported by the King Abdullah University of Science and Technology (KAUST) through the Office of Sponsored Research (OSR) [Grant Number OSR-2020-CRG8-4048]. We acknowledge computing resources on the supercomputer SHAHEEN granted by the KAUST Supercomputing Lab.

	\subsection*{Methods}
	
	The numerical calculations are performed using the time dependent quantum transport package KWANT \cite{Weston2016,Gaury2014,Kloss2021} where the stationary scattering properties are obtained from the dc transport package KWANT \cite{Groth2014}. 
	
	We consider a two dimensional tight binding model including both exchange and Rashba spin orbit coupling. The underlying Hamiltonian is given by 
	\begin{equation}
	\mathcal{H}(t) = \mathcal{H}_0(t) + \mathcal{H}_{\rm{R}},
	\end{equation}
	where,
	\begin{equation}
	\mathcal{H}_0 (t) = \sum_{r}\hat{c}^{\dagger}_r ( \hat{{\bm\sigma}}\cdot\mathbf{m}(r, t))\hat{c}_r -\gamma \sum_{\langle r,{r'}\rangle} \left(\hat{c}^{\dagger}_r \hat{c}_{r'} + \rm{h}.\rm{c}\right),
	\end{equation}
	
	with $\rm{h}.\rm{c}$ standing for hermitian conjugate, $\mathbf{m}(r, t)$ being the time dependent magnetization which also depend on position in the case of the antiferromagnetic dynamics. 
	The operator $\hat{{\bm\sigma}}$ represents the vector of Pauli matrices and $\gamma$ the tight binding hoping energy.
	The operators $\hat{c}^{\dagger}_r$ and $\hat{c}_r$ are respectively the creation and annihilation operators at position $r$ given by the coordinates $x$ and $y$. 
	
	The Rashba Hamiltonian is given by
	\begin{equation}
	\mathcal{H}_{\rm{R}} =  i \gamma \left(\frac{\alpha_{\rm{R}}}{2}\right) \sum_{r} \hat{c}^{\dagger}_{x, y} (\hat{{\sigma}}_y \hat{c}_{x+1, y} -\hat{{\sigma}}_x \hat{c}_{x, y+1} ) + \rm{h}.\rm{c},
	\end{equation}
	with $\alpha_{\rm{R}}$ the Rashba parameter.
	The system is connected to two transversal normal leads that allows for probing the pumped currents.
	
	To compute the non-equilibrium charge current, the stationary scattering modes of the tight binding system at different energies are obtained using KWANT. Subsequently, they are evolved forward in time according to the  time dependent Schr{\"o}dinger equation,
	\begin{equation}
	i\hbar \partial_t \Psi_{\eta m \varepsilon}(t) = \mathcal{H}(t) \Psi_{\eta m\varepsilon}(t),
	\end{equation}
	here, $\eta$ stands for the lead underlying the state, $m$  for the mode index and $\varepsilon$ for energy.
	Furthermore, the time dependent current at temperature T is obtained as 
	\begin{equation}
	\label{eq:it}
	{\rm {I} }(t) = \sum_{\eta m} \int \frac{d \varepsilon}{2 \pi} f (\varepsilon, \rm{T})\Psi_{\eta m\varepsilon}^{\dagger}(t) \Psi_{\eta m\varepsilon}(t),
	\end{equation}
	with $f(\varepsilon, \rm{T})$ being the Fermi distribution function at energy $\varepsilon$ and temperature T.
	
	All the numerical calculations are performed at a chemical potential of $100$ meV. Therefore, the calculated currents are summed from the bottom of the bands to the considered energy. The sd exchange coupling and the temperature of the system are set to $J=500$ meV and $T=0$ K respectively, except otherwise stated. To obtain the charge currents  in the frequency domain, a discrete fast Fourier transform of the time dependent signal is performed. In the data presented throughout the text $\rm{I}_{\omega}$ represents the absolute value of the normalized Fourier transform of $\rm{I}(t)$ given by Eq. \ref{eq:it}.

	\bibliography{refultra}

\begin{thebibliography}{33}%
\makeatletter
\providecommand \@ifxundefined [1]{%
 \@ifx{#1\undefined}
}%
\providecommand \@ifnum [1]{%
 \ifnum #1\expandafter \@firstoftwo
 \else \expandafter \@secondoftwo
 \fi
}%
\providecommand \@ifx [1]{%
 \ifx #1\expandafter \@firstoftwo
 \else \expandafter \@secondoftwo
 \fi
}%
\providecommand \natexlab [1]{#1}%
\providecommand \enquote  [1]{``#1''}%
\providecommand \bibnamefont  [1]{#1}%
\providecommand \bibfnamefont [1]{#1}%
\providecommand \citenamefont [1]{#1}%
\providecommand \href@noop [0]{\@secondoftwo}%
\providecommand \href [0]{\begingroup \@sanitize@url \@href}%
\providecommand \@href[1]{\@@startlink{#1}\@@href}%
\providecommand \@@href[1]{\endgroup#1\@@endlink}%
\providecommand \@sanitize@url [0]{\catcode `\\12\catcode `\$12\catcode
  `\&12\catcode `\#12\catcode `\^12\catcode `\_12\catcode `\%12\relax}%
\providecommand \@@startlink[1]{}%
\providecommand \@@endlink[0]{}%
\providecommand \url  [0]{\begingroup\@sanitize@url \@url }%
\providecommand \@url [1]{\endgroup\@href {#1}{\urlprefix }}%
\providecommand \urlprefix  [0]{URL }%
\providecommand \Eprint [0]{\href }%
\providecommand \doibase [0]{http://dx.doi.org/}%
\providecommand \selectlanguage [0]{\@gobble}%
\providecommand \bibinfo  [0]{\@secondoftwo}%
\providecommand \bibfield  [0]{\@secondoftwo}%
\providecommand \translation [1]{[#1]}%
\providecommand \BibitemOpen [0]{}%
\providecommand \bibitemStop [0]{}%
\providecommand \bibitemNoStop [0]{.\EOS\space}%
\providecommand \EOS [0]{\spacefactor3000\relax}%
\providecommand \BibitemShut  [1]{\csname bibitem#1\endcsname}%
\let\auto@bib@innerbib\@empty
\bibitem [{\citenamefont {Sinova}\ \emph {et~al.}(2015)\citenamefont {Sinova},
  \citenamefont {Valenzuela}, \citenamefont {Wunderlich}, \citenamefont
  {Back},\ and\ \citenamefont {Jungwirth}}]{Sinova2015}%
  \BibitemOpen
  \bibfield  {author} {\bibinfo {author} {\bibfnamefont {Jairo}\ \bibnamefont
  {Sinova}}, \bibinfo {author} {\bibfnamefont {Sergio~O.}\ \bibnamefont
  {Valenzuela}}, \bibinfo {author} {\bibfnamefont {J.}~\bibnamefont
  {Wunderlich}}, \bibinfo {author} {\bibfnamefont {C.~H.}\ \bibnamefont
  {Back}}, \ and\ \bibinfo {author} {\bibfnamefont {T.}~\bibnamefont
  {Jungwirth}},\ }\bibfield  {title} {\enquote {\bibinfo {title} {{Spin Hall
  effect}},}\ }\href@noop {} {\bibfield  {journal} {\bibinfo  {journal} {Review
  of Modern Physics}\ }\textbf {\bibinfo {volume} {87}},\ \bibinfo {pages}
  {1213} (\bibinfo {year} {2015})}\BibitemShut {NoStop}%
\bibitem [{\citenamefont {Edelstein}(1990)}]{Edelstein1990}%
  \BibitemOpen
  \bibfield  {author} {\bibinfo {author} {\bibfnamefont {V.M.}\ \bibnamefont
  {Edelstein}},\ }\bibfield  {title} {\enquote {\bibinfo {title} {Spin
  polarization of conduction electrons induced by electric current in
  two-dimensional asymmetric electron systems},}\ }\href {\doibase
  https://doi.org/10.1016/0038-1098(90)90963-C} {\bibfield  {journal} {\bibinfo
   {journal} {Solid State Communications}\ }\textbf {\bibinfo {volume} {73}},\
  \bibinfo {pages} {233 -- 235} (\bibinfo {year} {1990})}\BibitemShut {NoStop}%
\bibitem [{\citenamefont {Kampfrath}\ \emph {et~al.}(2013)\citenamefont
  {Kampfrath}, \citenamefont {Battiato}, \citenamefont {Maldonado},
  \citenamefont {Eilers}, \citenamefont {N{\"{o}}tzold}, \citenamefont
  {M{\"{a}}hrlein}, \citenamefont {Zbarsky}, \citenamefont {Freimuth},
  \citenamefont {Mokrousov}, \citenamefont {Bl{\"{u}}gel}, \citenamefont
  {Wolf}, \citenamefont {Radu}, \citenamefont {Oppeneer},\ and\ \citenamefont
  {M{\"{u}}nzenberg}}]{Kampfrath2013}%
  \BibitemOpen
  \bibfield  {author} {\bibinfo {author} {\bibfnamefont {Tobias}\ \bibnamefont
  {Kampfrath}}, \bibinfo {author} {\bibfnamefont {M}~\bibnamefont {Battiato}},
  \bibinfo {author} {\bibfnamefont {P}~\bibnamefont {Maldonado}}, \bibinfo
  {author} {\bibfnamefont {G}~\bibnamefont {Eilers}}, \bibinfo {author}
  {\bibfnamefont {J}~\bibnamefont {N{\"{o}}tzold}}, \bibinfo {author}
  {\bibfnamefont {S}~\bibnamefont {M{\"{a}}hrlein}}, \bibinfo {author}
  {\bibfnamefont {V}~\bibnamefont {Zbarsky}}, \bibinfo {author} {\bibfnamefont
  {Frank}\ \bibnamefont {Freimuth}}, \bibinfo {author} {\bibfnamefont {Yuriy}\
  \bibnamefont {Mokrousov}}, \bibinfo {author} {\bibfnamefont {Stefan}\
  \bibnamefont {Bl{\"{u}}gel}}, \bibinfo {author} {\bibfnamefont
  {M}~\bibnamefont {Wolf}}, \bibinfo {author} {\bibfnamefont {I}~\bibnamefont
  {Radu}}, \bibinfo {author} {\bibfnamefont {P~M}\ \bibnamefont {Oppeneer}}, \
  and\ \bibinfo {author} {\bibfnamefont {M.}~\bibnamefont {M{\"{u}}nzenberg}},\
  }\bibfield  {title} {\enquote {\bibinfo {title} {{Terahertz spin current
  pulses controlled by magnetic heterostructures.}}}\ }\href {\doibase
  10.1038/nnano.2013.43} {\bibfield  {journal} {\bibinfo  {journal} {Nature
  Nanotechnology}\ }\textbf {\bibinfo {volume} {8}},\ \bibinfo {pages} {256}
  (\bibinfo {year} {2013})}\BibitemShut {NoStop}%
\bibitem [{\citenamefont {Huisman}\ \emph {et~al.}(2016)\citenamefont
  {Huisman}, \citenamefont {Mikhaylovskiy}, \citenamefont {Costa},
  \citenamefont {Freimuth}, \citenamefont {Paz}, \citenamefont {Ventura},
  \citenamefont {Freitas}, \citenamefont {Bl{\"{u}}gel}, \citenamefont
  {Mokrousov}, \citenamefont {Rasing},\ and\ \citenamefont
  {Kimel}}]{Huisman2016}%
  \BibitemOpen
  \bibfield  {author} {\bibinfo {author} {\bibfnamefont {T~J}\ \bibnamefont
  {Huisman}}, \bibinfo {author} {\bibfnamefont {R~V}\ \bibnamefont
  {Mikhaylovskiy}}, \bibinfo {author} {\bibfnamefont {J~D}\ \bibnamefont
  {Costa}}, \bibinfo {author} {\bibfnamefont {F}~\bibnamefont {Freimuth}},
  \bibinfo {author} {\bibfnamefont {E}~\bibnamefont {Paz}}, \bibinfo {author}
  {\bibfnamefont {J}~\bibnamefont {Ventura}}, \bibinfo {author} {\bibfnamefont
  {P~P}\ \bibnamefont {Freitas}}, \bibinfo {author} {\bibfnamefont
  {S}~\bibnamefont {Bl{\"{u}}gel}}, \bibinfo {author} {\bibfnamefont
  {Y}~\bibnamefont {Mokrousov}}, \bibinfo {author} {\bibfnamefont {Th.}\
  \bibnamefont {Rasing}}, \ and\ \bibinfo {author} {\bibfnamefont {A~V}\
  \bibnamefont {Kimel}},\ }\bibfield  {title} {\enquote {\bibinfo {title}
  {{Femtosecond control of electric currents in metallic ferromagnetic
  heterostructures}},}\ }\href {\doibase 10.1038/nnano.2015.331} {\bibfield
  {journal} {\bibinfo  {journal} {Nature Nanotechnology}\ }\textbf {\bibinfo
  {volume} {11}},\ \bibinfo {pages} {455} (\bibinfo {year} {2016})},\ \Eprint
  {http://arxiv.org/abs/1505.02970} {arXiv:1505.02970} \BibitemShut {NoStop}%
\bibitem [{\citenamefont {Brataas}\ \emph {et~al.}(2002)\citenamefont
  {Brataas}, \citenamefont {Tserkovnyak}, \citenamefont {Bauer},\ and\
  \citenamefont {Halperin}}]{Brataas2002}%
  \BibitemOpen
  \bibfield  {author} {\bibinfo {author} {\bibfnamefont {Arne}\ \bibnamefont
  {Brataas}}, \bibinfo {author} {\bibfnamefont {Yaroslav}\ \bibnamefont
  {Tserkovnyak}}, \bibinfo {author} {\bibfnamefont {G.~E.~W.}\ \bibnamefont
  {Bauer}}, \ and\ \bibinfo {author} {\bibfnamefont {Bertrand}\ \bibnamefont
  {Halperin}},\ }\bibfield  {title} {\enquote {\bibinfo {title} {{Spin battery
  operated by ferromagnetic resonance}},}\ }\href {\doibase
  10.1103/PhysRevB.66.060404} {\bibfield  {journal} {\bibinfo  {journal}
  {Physical Review B}\ }\textbf {\bibinfo {volume} {66}},\ \bibinfo {pages}
  {060404} (\bibinfo {year} {2002})}\BibitemShut {NoStop}%
\bibitem [{\citenamefont {Saitoh}\ \emph {et~al.}(2006)\citenamefont {Saitoh},
  \citenamefont {Ueda}, \citenamefont {Miyajima},\ and\ \citenamefont
  {Tatara}}]{Saitoh2006}%
  \BibitemOpen
  \bibfield  {author} {\bibinfo {author} {\bibfnamefont {E.}~\bibnamefont
  {Saitoh}}, \bibinfo {author} {\bibfnamefont {M.}~\bibnamefont {Ueda}},
  \bibinfo {author} {\bibfnamefont {H.}~\bibnamefont {Miyajima}}, \ and\
  \bibinfo {author} {\bibfnamefont {G.}~\bibnamefont {Tatara}},\ }\bibfield
  {title} {\enquote {\bibinfo {title} {{Conversion of spin current into charge
  current at room temperature: Inverse spin-Hall effect}},}\ }\href {\doibase
  10.1063/1.2199473} {\bibfield  {journal} {\bibinfo  {journal} {Applied
  Physics Letters}\ }\textbf {\bibinfo {volume} {88}},\ \bibinfo {pages}
  {182509} (\bibinfo {year} {2006})}\BibitemShut {NoStop}%
\bibitem [{\citenamefont {Corkum}(1993)}]{Corkum1993}%
  \BibitemOpen
  \bibfield  {author} {\bibinfo {author} {\bibfnamefont {P.~B.}\ \bibnamefont
  {Corkum}},\ }\bibfield  {title} {\enquote {\bibinfo {title} {Plasma
  perspective on strong field multiphoton ionization},}\ }\href {\doibase
  10.1103/PhysRevLett.71.1994} {\bibfield  {journal} {\bibinfo  {journal}
  {Phys. Rev. Lett.}\ }\textbf {\bibinfo {volume} {71}},\ \bibinfo {pages}
  {1994--1997} (\bibinfo {year} {1993})}\BibitemShut {NoStop}%
\bibitem [{\citenamefont {Corkum}\ and\ \citenamefont
  {Krausz}(2007)}]{Corkum2007}%
  \BibitemOpen
  \bibfield  {author} {\bibinfo {author} {\bibfnamefont {P.~B.}\ \bibnamefont
  {Corkum}}\ and\ \bibinfo {author} {\bibfnamefont {Ferenc}\ \bibnamefont
  {Krausz}},\ }\bibfield  {title} {\enquote {\bibinfo {title} {Attosecond
  science},}\ }\href {\doibase 10.1038/nphys620} {\bibfield  {journal}
  {\bibinfo  {journal} {Nature Physics}\ }\textbf {\bibinfo {volume} {3}},\
  \bibinfo {pages} {381--387} (\bibinfo {year} {2007})}\BibitemShut {NoStop}%
\bibitem [{\citenamefont {Midorikawa}(2011)}]{Midorikawa2011}%
  \BibitemOpen
  \bibfield  {author} {\bibinfo {author} {\bibfnamefont {Katsumi}\ \bibnamefont
  {Midorikawa}},\ }\bibfield  {title} {\enquote {\bibinfo {title} {Ultrafast
  dynamic imaging},}\ }\href {\doibase 10.1038/nphoton.2011.265} {\bibfield
  {journal} {\bibinfo  {journal} {Nature Photonics}\ }\textbf {\bibinfo
  {volume} {5}},\ \bibinfo {pages} {640--641} (\bibinfo {year}
  {2011})}\BibitemShut {NoStop}%
\bibitem [{\citenamefont {Shiomi}\ \emph {et~al.}(2014)\citenamefont {Shiomi},
  \citenamefont {Nomura}, \citenamefont {Kajiwara}, \citenamefont {Eto},
  \citenamefont {Novak}, \citenamefont {Segawa}, \citenamefont {Ando},\ and\
  \citenamefont {Saitoh}}]{Shiomi2014}%
  \BibitemOpen
  \bibfield  {author} {\bibinfo {author} {\bibfnamefont {Y.}~\bibnamefont
  {Shiomi}}, \bibinfo {author} {\bibfnamefont {K.}~\bibnamefont {Nomura}},
  \bibinfo {author} {\bibfnamefont {Y.}~\bibnamefont {Kajiwara}}, \bibinfo
  {author} {\bibfnamefont {K.}~\bibnamefont {Eto}}, \bibinfo {author}
  {\bibfnamefont {M.}~\bibnamefont {Novak}}, \bibinfo {author} {\bibfnamefont
  {Kouji}\ \bibnamefont {Segawa}}, \bibinfo {author} {\bibfnamefont {Yoichi}\
  \bibnamefont {Ando}}, \ and\ \bibinfo {author} {\bibfnamefont
  {E.}~\bibnamefont {Saitoh}},\ }\bibfield  {title} {\enquote {\bibinfo {title}
  {{Spin-Electricity Conversion Induced by Spin Injection into Topological
  Insulators}},}\ }\href {\doibase 10.1103/PhysRevLett.113.196601} {\bibfield
  {journal} {\bibinfo  {journal} {Physical Review Letters}\ }\textbf {\bibinfo
  {volume} {113}},\ \bibinfo {pages} {196601} (\bibinfo {year}
  {2014})}\BibitemShut {NoStop}%
\bibitem [{\citenamefont {Caviglia}\ \emph {et~al.}(2010)\citenamefont
  {Caviglia}, \citenamefont {Gabay}, \citenamefont {Gariglio}, \citenamefont
  {Reyren}, \citenamefont {Cancellieri},\ and\ \citenamefont
  {Triscone}}]{Caviglia2010}%
  \BibitemOpen
  \bibfield  {author} {\bibinfo {author} {\bibfnamefont {A.~D.}\ \bibnamefont
  {Caviglia}}, \bibinfo {author} {\bibfnamefont {M.}~\bibnamefont {Gabay}},
  \bibinfo {author} {\bibfnamefont {S.}~\bibnamefont {Gariglio}}, \bibinfo
  {author} {\bibfnamefont {N.}~\bibnamefont {Reyren}}, \bibinfo {author}
  {\bibfnamefont {C.}~\bibnamefont {Cancellieri}}, \ and\ \bibinfo {author}
  {\bibfnamefont {J.-M.}\ \bibnamefont {Triscone}},\ }\bibfield  {title}
  {\enquote {\bibinfo {title} {{Tunable Rashba Spin-Orbit Interaction at Oxide
  Interfaces}},}\ }\href {\doibase 10.1103/PhysRevLett.104.126803} {\bibfield
  {journal} {\bibinfo  {journal} {Phys. Rev. Lett.}\ }\textbf {\bibinfo
  {volume} {104}},\ \bibinfo {pages} {126803} (\bibinfo {year}
  {2010})}\BibitemShut {NoStop}%
\bibitem [{\citenamefont {Vaz}\ \emph {et~al.}(2019)\citenamefont {Vaz},
  \citenamefont {No{\"{e}}l}, \citenamefont {Johansson}, \citenamefont
  {G{\"{o}}bel}, \citenamefont {Bruno}, \citenamefont {Singh}, \citenamefont
  {Mckeown-walker}, \citenamefont {Trier}, \citenamefont {Vicente-arche},
  \citenamefont {Sander}, \citenamefont {Valencia}, \citenamefont {Bruneel},
  \citenamefont {Vivek}, \citenamefont {Gabay}, \citenamefont {Bergeal},
  \citenamefont {Baumberger}, \citenamefont {Okuno}, \citenamefont
  {Barth{\'{e}}l{\'{e}}my}, \citenamefont {Fert}, \citenamefont {Vila},
  \citenamefont {Mertig}, \citenamefont {Attan{\'{e}}},\ and\ \citenamefont
  {Bibes}}]{Vaz2019}%
  \BibitemOpen
  \bibfield  {author} {\bibinfo {author} {\bibfnamefont {Diogo~C}\ \bibnamefont
  {Vaz}}, \bibinfo {author} {\bibfnamefont {Paul}\ \bibnamefont {No{\"{e}}l}},
  \bibinfo {author} {\bibfnamefont {Annika}\ \bibnamefont {Johansson}},
  \bibinfo {author} {\bibfnamefont {B{\"{o}}rge}\ \bibnamefont {G{\"{o}}bel}},
  \bibinfo {author} {\bibfnamefont {Flavio~Y}\ \bibnamefont {Bruno}}, \bibinfo
  {author} {\bibfnamefont {Gyanendra}\ \bibnamefont {Singh}}, \bibinfo {author}
  {\bibfnamefont {Siobhan}\ \bibnamefont {Mckeown-walker}}, \bibinfo {author}
  {\bibfnamefont {Felix}\ \bibnamefont {Trier}}, \bibinfo {author}
  {\bibfnamefont {Luis~M}\ \bibnamefont {Vicente-arche}}, \bibinfo {author}
  {\bibfnamefont {Anke}\ \bibnamefont {Sander}}, \bibinfo {author}
  {\bibfnamefont {Sergio}\ \bibnamefont {Valencia}}, \bibinfo {author}
  {\bibfnamefont {Pierre}\ \bibnamefont {Bruneel}}, \bibinfo {author}
  {\bibfnamefont {Manali}\ \bibnamefont {Vivek}}, \bibinfo {author}
  {\bibfnamefont {Marc}\ \bibnamefont {Gabay}}, \bibinfo {author}
  {\bibfnamefont {Nicolas}\ \bibnamefont {Bergeal}}, \bibinfo {author}
  {\bibfnamefont {Felix}\ \bibnamefont {Baumberger}}, \bibinfo {author}
  {\bibfnamefont {Hanako}\ \bibnamefont {Okuno}}, \bibinfo {author}
  {\bibfnamefont {Agn{\`{e}}s}\ \bibnamefont {Barth{\'{e}}l{\'{e}}my}},
  \bibinfo {author} {\bibfnamefont {Albert}\ \bibnamefont {Fert}}, \bibinfo
  {author} {\bibfnamefont {Laurent}\ \bibnamefont {Vila}}, \bibinfo {author}
  {\bibfnamefont {Ingrid}\ \bibnamefont {Mertig}}, \bibinfo {author}
  {\bibfnamefont {Jean-philippe}\ \bibnamefont {Attan{\'{e}}}}, \ and\ \bibinfo
  {author} {\bibfnamefont {Manuel}\ \bibnamefont {Bibes}},\ }\bibfield  {title}
  {\enquote {\bibinfo {title} {{Mapping spin–charge conversion to the band
  structure in a topological oxide two-dimensional electron gas}},}\ }\href
  {\doibase 10.1038/s41563-019-0467-4} {\bibfield  {journal} {\bibinfo
  {journal} {Nature Materials}\ }\textbf {\bibinfo {volume} {18}},\ \bibinfo
  {pages} {1187} (\bibinfo {year} {2019})}\BibitemShut {NoStop}%
\bibitem [{\citenamefont {Wei}\ \emph {et~al.}(2014)\citenamefont {Wei},
  \citenamefont {Obstbaum}, \citenamefont {Ribow}, \citenamefont {Back},\ and\
  \citenamefont {Woltersdorf}}]{Wei2014}%
  \BibitemOpen
  \bibfield  {author} {\bibinfo {author} {\bibfnamefont {Dahai}\ \bibnamefont
  {Wei}}, \bibinfo {author} {\bibfnamefont {Martin}\ \bibnamefont {Obstbaum}},
  \bibinfo {author} {\bibfnamefont {Mirko}\ \bibnamefont {Ribow}}, \bibinfo
  {author} {\bibfnamefont {C.~H.}\ \bibnamefont {Back}}, \ and\ \bibinfo
  {author} {\bibfnamefont {Georg}\ \bibnamefont {Woltersdorf}},\ }\bibfield
  {title} {\enquote {\bibinfo {title} {{Spin Hall voltages from a.c. and d.c.
  spin currents.}}}\ }\href {\doibase 10.1038/ncomms4768} {\bibfield  {journal}
  {\bibinfo  {journal} {Nature Communications}\ }\textbf {\bibinfo {volume}
  {5}},\ \bibinfo {pages} {3768} (\bibinfo {year} {2014})}\BibitemShut
  {NoStop}%
\bibitem [{\citenamefont {Rojas-S{\'{a}}nchez}\ \emph
  {et~al.}(2013)\citenamefont {Rojas-S{\'{a}}nchez}, \citenamefont {Vila},
  \citenamefont {Desfonds}, \citenamefont {Gambarelli}, \citenamefont {Attane},
  \citenamefont {{De Teresa}}, \citenamefont {Mag{\'{e}}n},\ and\ \citenamefont
  {Fert}}]{Rojas-Sanchez2013b}%
  \BibitemOpen
  \bibfield  {author} {\bibinfo {author} {\bibfnamefont {J~C}\ \bibnamefont
  {Rojas-S{\'{a}}nchez}}, \bibinfo {author} {\bibfnamefont {L}~\bibnamefont
  {Vila}}, \bibinfo {author} {\bibfnamefont {G}~\bibnamefont {Desfonds}},
  \bibinfo {author} {\bibfnamefont {S}~\bibnamefont {Gambarelli}}, \bibinfo
  {author} {\bibfnamefont {J.-P.}\ \bibnamefont {Attane}}, \bibinfo {author}
  {\bibfnamefont {J~M}\ \bibnamefont {{De Teresa}}}, \bibinfo {author}
  {\bibfnamefont {C}~\bibnamefont {Mag{\'{e}}n}}, \ and\ \bibinfo {author}
  {\bibfnamefont {A.}~\bibnamefont {Fert}},\ }\bibfield  {title} {\enquote
  {\bibinfo {title} {{Spin-to-charge conversion using Rashba coupling at the
  interface between non-magnetic materials.}}}\ }\href {\doibase
  10.1038/ncomms3944} {\bibfield  {journal} {\bibinfo  {journal} {Nature
  Communications}\ }\textbf {\bibinfo {volume} {4}},\ \bibinfo {pages} {2944}
  (\bibinfo {year} {2013})}\BibitemShut {NoStop}%
\bibitem [{\citenamefont {Weston}\ and\ \citenamefont
  {Waintal}(2016)}]{Weston2016}%
  \BibitemOpen
  \bibfield  {author} {\bibinfo {author} {\bibfnamefont {Joseph}\ \bibnamefont
  {Weston}}\ and\ \bibinfo {author} {\bibfnamefont {Xavier}\ \bibnamefont
  {Waintal}},\ }\bibfield  {title} {\enquote {\bibinfo {title} {{Linear-scaling
  source-sink algorithm for simulating time-resolved quantum transport and
  superconductivity}},}\ }\href {\doibase 10.1103/PhysRevB.93.134506}
  {\bibfield  {journal} {\bibinfo  {journal} {Physical Review B}\ }\textbf
  {\bibinfo {volume} {93}},\ \bibinfo {pages} {134506} (\bibinfo {year}
  {2016})}\BibitemShut {NoStop}%
\bibitem [{\citenamefont {Gaury}\ \emph {et~al.}(2014)\citenamefont {Gaury},
  \citenamefont {Weston}, \citenamefont {Santin}, \citenamefont {Houzet},
  \citenamefont {Groth},\ and\ \citenamefont {Waintal}}]{Gaury2014}%
  \BibitemOpen
  \bibfield  {author} {\bibinfo {author} {\bibfnamefont {Benoit}\ \bibnamefont
  {Gaury}}, \bibinfo {author} {\bibfnamefont {Joseph}\ \bibnamefont {Weston}},
  \bibinfo {author} {\bibfnamefont {Matthieu}\ \bibnamefont {Santin}}, \bibinfo
  {author} {\bibfnamefont {Manuel}\ \bibnamefont {Houzet}}, \bibinfo {author}
  {\bibfnamefont {Christoph}\ \bibnamefont {Groth}}, \ and\ \bibinfo {author}
  {\bibfnamefont {Xavier}\ \bibnamefont {Waintal}},\ }\href {\doibase
  10.1016/j.physrep.2013.09.001} {\enquote {\bibinfo {title} {{Numerical
  simulations of time-resolved quantum electronics}},}\ } (\bibinfo {year}
  {2014})\BibitemShut {NoStop}%
\bibitem [{\citenamefont {Kloss}\ \emph {et~al.}(2021)\citenamefont {Kloss},
  \citenamefont {Weston}, \citenamefont {Gaury}, \citenamefont {Rossignol},
  \citenamefont {Groth},\ and\ \citenamefont {Waintal}}]{Kloss2021}%
  \BibitemOpen
  \bibfield  {author} {\bibinfo {author} {\bibfnamefont {Thomas}\ \bibnamefont
  {Kloss}}, \bibinfo {author} {\bibfnamefont {Joseph}\ \bibnamefont {Weston}},
  \bibinfo {author} {\bibfnamefont {Benoit}\ \bibnamefont {Gaury}}, \bibinfo
  {author} {\bibfnamefont {Benoit}\ \bibnamefont {Rossignol}}, \bibinfo
  {author} {\bibfnamefont {Christoph}\ \bibnamefont {Groth}}, \ and\ \bibinfo
  {author} {\bibfnamefont {Xavier}\ \bibnamefont {Waintal}},\ }\bibfield
  {title} {\enquote {\bibinfo {title} {Tkwant: a software package for
  time-dependent quantum transport},}\ }\href {\doibase
  10.1088/1367-2630/abddf7} {\bibfield  {journal} {\bibinfo  {journal} {New
  Journal of Physics}\ }\textbf {\bibinfo {volume} {23}},\ \bibinfo {pages}
  {023025} (\bibinfo {year} {2021})}\BibitemShut {NoStop}%
\bibitem [{Sup()}]{SuppMat}%
  \BibitemOpen
  \href@noop {} {\enquote {\bibinfo {title} {{Supplemental Materials}},}\
  }\BibitemShut {NoStop}%
\bibitem [{\citenamefont {Cheng}\ \emph {et~al.}(2014)\citenamefont {Cheng},
  \citenamefont {Xiao}, \citenamefont {Niu},\ and\ \citenamefont
  {Brataas}}]{Cheng2014c}%
  \BibitemOpen
  \bibfield  {author} {\bibinfo {author} {\bibfnamefont {Ran}\ \bibnamefont
  {Cheng}}, \bibinfo {author} {\bibfnamefont {Jiang}\ \bibnamefont {Xiao}},
  \bibinfo {author} {\bibfnamefont {Qian}\ \bibnamefont {Niu}}, \ and\ \bibinfo
  {author} {\bibfnamefont {Arne}\ \bibnamefont {Brataas}},\ }\bibfield  {title}
  {\enquote {\bibinfo {title} {{Spin pumping and spin-transfer torques in
  antiferromagnets}},}\ }\href {\doibase 10.1103/PhysRevLett.113.057601}
  {\bibfield  {journal} {\bibinfo  {journal} {Physical Review Letters}\
  }\textbf {\bibinfo {volume} {113}},\ \bibinfo {pages} {057601} (\bibinfo
  {year} {2014})}\BibitemShut {NoStop}%
\bibitem [{\citenamefont {Keffer}\ and\ \citenamefont
  {Kittel}(1952)}]{Keffer1952}%
  \BibitemOpen
  \bibfield  {author} {\bibinfo {author} {\bibfnamefont {F.}~\bibnamefont
  {Keffer}}\ and\ \bibinfo {author} {\bibfnamefont {C.}~\bibnamefont
  {Kittel}},\ }\bibfield  {title} {\enquote {\bibinfo {title} {{Theory of
  Antiferromagnetic Resonance}},}\ }\href@noop {} {\bibfield  {journal}
  {\bibinfo  {journal} {Phy. Rev.}\ }\textbf {\bibinfo {volume} {85}},\
  \bibinfo {pages} {329} (\bibinfo {year} {1952})}\BibitemShut {NoStop}%
\bibitem [{\citenamefont {Ast}\ \emph {et~al.}(2007)\citenamefont {Ast},
  \citenamefont {Henk}, \citenamefont {Ernst}, \citenamefont {Moreschini},
  \citenamefont {Falub}, \citenamefont {Pacil\'e}, \citenamefont {Bruno},
  \citenamefont {Kern},\ and\ \citenamefont {Grioni}}]{Ast2007}%
  \BibitemOpen
  \bibfield  {author} {\bibinfo {author} {\bibfnamefont {Christian~R.}\
  \bibnamefont {Ast}}, \bibinfo {author} {\bibfnamefont {J\"urgen}\
  \bibnamefont {Henk}}, \bibinfo {author} {\bibfnamefont {Arthur}\ \bibnamefont
  {Ernst}}, \bibinfo {author} {\bibfnamefont {Luca}\ \bibnamefont
  {Moreschini}}, \bibinfo {author} {\bibfnamefont {Mihaela~C.}\ \bibnamefont
  {Falub}}, \bibinfo {author} {\bibfnamefont {Daniela}\ \bibnamefont
  {Pacil\'e}}, \bibinfo {author} {\bibfnamefont {Patrick}\ \bibnamefont
  {Bruno}}, \bibinfo {author} {\bibfnamefont {Klaus}\ \bibnamefont {Kern}}, \
  and\ \bibinfo {author} {\bibfnamefont {Marco}\ \bibnamefont {Grioni}},\
  }\bibfield  {title} {\enquote {\bibinfo {title} {Giant spin splitting through
  surface alloying},}\ }\href {\doibase 10.1103/PhysRevLett.98.186807}
  {\bibfield  {journal} {\bibinfo  {journal} {Phys. Rev. Lett.}\ }\textbf
  {\bibinfo {volume} {98}},\ \bibinfo {pages} {186807} (\bibinfo {year}
  {2007})}\BibitemShut {NoStop}%
\bibitem [{\citenamefont {Zhang}\ \emph {et~al.}(2009)\citenamefont {Zhang},
  \citenamefont {Liu}, \citenamefont {Qi}, \citenamefont {Dai}, \citenamefont
  {Fang},\ and\ \citenamefont {Zhang}}]{Zhang2009}%
  \BibitemOpen
  \bibfield  {author} {\bibinfo {author} {\bibfnamefont {Haijun}\ \bibnamefont
  {Zhang}}, \bibinfo {author} {\bibfnamefont {Chao-Xing}\ \bibnamefont {Liu}},
  \bibinfo {author} {\bibfnamefont {Xiao-Liang}\ \bibnamefont {Qi}}, \bibinfo
  {author} {\bibfnamefont {Xi}~\bibnamefont {Dai}}, \bibinfo {author}
  {\bibfnamefont {Zhong}\ \bibnamefont {Fang}}, \ and\ \bibinfo {author}
  {\bibfnamefont {Shou-Cheng}\ \bibnamefont {Zhang}},\ }\bibfield  {title}
  {\enquote {\bibinfo {title} {Topological insulators in bi2se3, bi2te3 and
  sb2te3 with a single dirac cone on the surface},}\ }\href {\doibase
  10.1038/nphys1270} {\bibfield  {journal} {\bibinfo  {journal} {Nature
  Physics}\ }\textbf {\bibinfo {volume} {5}},\ \bibinfo {pages} {438--442}
  (\bibinfo {year} {2009})}\BibitemShut {NoStop}%
\bibitem [{\citenamefont {Hafez}\ \emph {et~al.}(2018)\citenamefont {Hafez},
  \citenamefont {Kovalev}, \citenamefont {Deinert}, \citenamefont {Mics},
  \citenamefont {Green}, \citenamefont {Awari}, \citenamefont {Chen},
  \citenamefont {Germanskiy}, \citenamefont {Lehnert}, \citenamefont
  {Teichert}, \citenamefont {Wang}, \citenamefont {Tielrooij}, \citenamefont
  {Liu}, \citenamefont {Chen}, \citenamefont {Narita}, \citenamefont
  {M{\"{u}}llen}, \citenamefont {Bonn}, \citenamefont {Gensch},\ and\
  \citenamefont {Turchinovich}}]{Hafez2018}%
  \BibitemOpen
  \bibfield  {author} {\bibinfo {author} {\bibfnamefont {Hassan~A}\
  \bibnamefont {Hafez}}, \bibinfo {author} {\bibfnamefont {Sergey}\
  \bibnamefont {Kovalev}}, \bibinfo {author} {\bibfnamefont {Jan-christoph}\
  \bibnamefont {Deinert}}, \bibinfo {author} {\bibfnamefont {Zolt{\'{a}}n}\
  \bibnamefont {Mics}}, \bibinfo {author} {\bibfnamefont {Bertram}\
  \bibnamefont {Green}}, \bibinfo {author} {\bibfnamefont {Nilesh}\
  \bibnamefont {Awari}}, \bibinfo {author} {\bibfnamefont {Min}\ \bibnamefont
  {Chen}}, \bibinfo {author} {\bibfnamefont {Semyon}\ \bibnamefont
  {Germanskiy}}, \bibinfo {author} {\bibfnamefont {Ulf}\ \bibnamefont
  {Lehnert}}, \bibinfo {author} {\bibfnamefont {Jochen}\ \bibnamefont
  {Teichert}}, \bibinfo {author} {\bibfnamefont {Zhe}\ \bibnamefont {Wang}},
  \bibinfo {author} {\bibfnamefont {Klaas-jan}\ \bibnamefont {Tielrooij}},
  \bibinfo {author} {\bibfnamefont {Zhaoyang}\ \bibnamefont {Liu}}, \bibinfo
  {author} {\bibfnamefont {Zongping}\ \bibnamefont {Chen}}, \bibinfo {author}
  {\bibfnamefont {Akimitsu}\ \bibnamefont {Narita}}, \bibinfo {author}
  {\bibfnamefont {Klaus}\ \bibnamefont {M{\"{u}}llen}}, \bibinfo {author}
  {\bibfnamefont {Mischa}\ \bibnamefont {Bonn}}, \bibinfo {author}
  {\bibfnamefont {Michael}\ \bibnamefont {Gensch}}, \ and\ \bibinfo {author}
  {\bibfnamefont {Dmitry}\ \bibnamefont {Turchinovich}},\ }\bibfield  {title}
  {\enquote {\bibinfo {title} {{Extremely efficient terahertz high-harmonic
  generation in graphene by hot Dirac fermions}},}\ }\href {\doibase
  10.1038/s41586-018-0508-1} {\bibfield  {journal} {\bibinfo  {journal}
  {Nature}\ }\textbf {\bibinfo {volume} {56}},\ \bibinfo {pages} {507}
  (\bibinfo {year} {2018})}\BibitemShut {NoStop}%
\bibitem [{\citenamefont {Ghimire}\ and\ \citenamefont
  {Reis}(2019)}]{Ghimire2019}%
  \BibitemOpen
  \bibfield  {author} {\bibinfo {author} {\bibfnamefont {Shambhu}\ \bibnamefont
  {Ghimire}}\ and\ \bibinfo {author} {\bibfnamefont {David~A.}\ \bibnamefont
  {Reis}},\ }\bibfield  {title} {\enquote {\bibinfo {title} {High-harmonic
  generation from solids},}\ }\href {\doibase 10.1038/s41567-018-0315-5}
  {\bibfield  {journal} {\bibinfo  {journal} {Nature Physics}\ }\textbf
  {\bibinfo {volume} {15}},\ \bibinfo {pages} {10--16} (\bibinfo {year}
  {2019})}\BibitemShut {NoStop}%
\bibitem [{\citenamefont {Luu}\ \emph {et~al.}(2015)\citenamefont {Luu},
  \citenamefont {Garg}, \citenamefont {Kruchinin}, \citenamefont {Moulet},
  \citenamefont {Hassan},\ and\ \citenamefont {Goulielmakis}}]{Luu2015}%
  \BibitemOpen
  \bibfield  {author} {\bibinfo {author} {\bibfnamefont {T.~T.}\ \bibnamefont
  {Luu}}, \bibinfo {author} {\bibfnamefont {M.}~\bibnamefont {Garg}}, \bibinfo
  {author} {\bibfnamefont {S.~Yu}\ \bibnamefont {Kruchinin}}, \bibinfo {author}
  {\bibfnamefont {A.}~\bibnamefont {Moulet}}, \bibinfo {author} {\bibfnamefont
  {M.~Th}\ \bibnamefont {Hassan}}, \ and\ \bibinfo {author} {\bibfnamefont
  {E.}~\bibnamefont {Goulielmakis}},\ }\bibfield  {title} {\enquote {\bibinfo
  {title} {Extreme ultraviolet high-harmonic spectroscopy of solids},}\ }\href
  {\doibase 10.1038/nature14456} {\bibfield  {journal} {\bibinfo  {journal}
  {Nature}\ }\textbf {\bibinfo {volume} {521}},\ \bibinfo {pages} {498--502}
  (\bibinfo {year} {2015})}\BibitemShut {NoStop}%
\bibitem [{\citenamefont {Wu}\ \emph {et~al.}(2015)\citenamefont {Wu},
  \citenamefont {Ghimire}, \citenamefont {Reis}, \citenamefont {Schafer},\ and\
  \citenamefont {Gaarde}}]{Wu2015}%
  \BibitemOpen
  \bibfield  {author} {\bibinfo {author} {\bibfnamefont {Mengxi}\ \bibnamefont
  {Wu}}, \bibinfo {author} {\bibfnamefont {Shambhu}\ \bibnamefont {Ghimire}},
  \bibinfo {author} {\bibfnamefont {David~A.}\ \bibnamefont {Reis}}, \bibinfo
  {author} {\bibfnamefont {Kenneth~J.}\ \bibnamefont {Schafer}}, \ and\
  \bibinfo {author} {\bibfnamefont {Mette~B.}\ \bibnamefont {Gaarde}},\
  }\bibfield  {title} {\enquote {\bibinfo {title} {High-harmonic generation
  from bloch electrons in solids},}\ }\href {\doibase
  10.1103/PhysRevA.91.043839} {\bibfield  {journal} {\bibinfo  {journal} {Phys.
  Rev. A}\ }\textbf {\bibinfo {volume} {91}},\ \bibinfo {pages} {043839}
  (\bibinfo {year} {2015})}\BibitemShut {NoStop}%
\bibitem [{\citenamefont {Jalabert}\ \emph {et~al.}(2010)\citenamefont
  {Jalabert}, \citenamefont {Szewc}, \citenamefont {Tomsovic},\ and\
  \citenamefont {Weinmann}}]{Jalabert2010}%
  \BibitemOpen
  \bibfield  {author} {\bibinfo {author} {\bibfnamefont {Rodolfo~A.}\
  \bibnamefont {Jalabert}}, \bibinfo {author} {\bibfnamefont {Wojciech}\
  \bibnamefont {Szewc}}, \bibinfo {author} {\bibfnamefont {Steven}\
  \bibnamefont {Tomsovic}}, \ and\ \bibinfo {author} {\bibfnamefont {Dietmar}\
  \bibnamefont {Weinmann}},\ }\bibfield  {title} {\enquote {\bibinfo {title}
  {What is measured in the scanning gate microscopy of a quantum point
  contact?}}\ }\href {\doibase 10.1103/PhysRevLett.105.166802} {\bibfield
  {journal} {\bibinfo  {journal} {Phys. Rev. Lett.}\ }\textbf {\bibinfo
  {volume} {105}},\ \bibinfo {pages} {166802} (\bibinfo {year}
  {2010})}\BibitemShut {NoStop}%
\bibitem [{\citenamefont {Gorini}\ \emph {et~al.}(2013)\citenamefont {Gorini},
  \citenamefont {Jalabert}, \citenamefont {Szewc}, \citenamefont {Tomsovic},\
  and\ \citenamefont {Weinmann}}]{Gorini2013}%
  \BibitemOpen
  \bibfield  {author} {\bibinfo {author} {\bibfnamefont {Cosimo}\ \bibnamefont
  {Gorini}}, \bibinfo {author} {\bibfnamefont {Rodolfo~A.}\ \bibnamefont
  {Jalabert}}, \bibinfo {author} {\bibfnamefont {Wojciech}\ \bibnamefont
  {Szewc}}, \bibinfo {author} {\bibfnamefont {Steven}\ \bibnamefont
  {Tomsovic}}, \ and\ \bibinfo {author} {\bibfnamefont {Dietmar}\ \bibnamefont
  {Weinmann}},\ }\bibfield  {title} {\enquote {\bibinfo {title} {Theory of
  scanning gate microscopy},}\ }\href {\doibase 10.1103/PhysRevB.88.035406}
  {\bibfield  {journal} {\bibinfo  {journal} {Phys. Rev. B}\ }\textbf {\bibinfo
  {volume} {88}},\ \bibinfo {pages} {035406} (\bibinfo {year}
  {2013})}\BibitemShut {NoStop}%
\bibitem [{\citenamefont {Manchon}\ \emph {et~al.}(2019)\citenamefont
  {Manchon}, \citenamefont {Zelezn{\'{y}}}, \citenamefont {Miron},
  \citenamefont {Jungwirth}, \citenamefont {Sinova}, \citenamefont {Thiaville},
  \citenamefont {Garello},\ and\ \citenamefont {Gambardella}}]{Manchon2019}%
  \BibitemOpen
  \bibfield  {author} {\bibinfo {author} {\bibfnamefont {A}~\bibnamefont
  {Manchon}}, \bibinfo {author} {\bibfnamefont {J.}~\bibnamefont
  {Zelezn{\'{y}}}}, \bibinfo {author} {\bibfnamefont {M.}~\bibnamefont
  {Miron}}, \bibinfo {author} {\bibfnamefont {Tom{\'{a}}{\v{s}}}\ \bibnamefont
  {Jungwirth}}, \bibinfo {author} {\bibfnamefont {Jairo}\ \bibnamefont
  {Sinova}}, \bibinfo {author} {\bibfnamefont {Andr{\'{e}}}\ \bibnamefont
  {Thiaville}}, \bibinfo {author} {\bibfnamefont {Kevin}\ \bibnamefont
  {Garello}}, \ and\ \bibinfo {author} {\bibfnamefont {Pietro}\ \bibnamefont
  {Gambardella}},\ }\bibfield  {title} {\enquote {\bibinfo {title}
  {{Current-induced spin-orbit torques in ferromagnetic and antiferromagnetic
  systems}},}\ }\href {\doibase 10.1103/RevModPhys.91.035004} {\bibfield
  {journal} {\bibinfo  {journal} {Review of Modern Physics}\ }\textbf {\bibinfo
  {volume} {91}},\ \bibinfo {pages} {035004} (\bibinfo {year}
  {2019})}\BibitemShut {NoStop}%
\bibitem [{\citenamefont {Lesne}\ \emph {et~al.}(2016)\citenamefont {Lesne},
  \citenamefont {Fu}, \citenamefont {Oyarzun}, \citenamefont
  {Rojas-S{\'{a}}nchez}, \citenamefont {Vaz}, \citenamefont {Naganuma},
  \citenamefont {Sicoli}, \citenamefont {Attan{\'{e}}}, \citenamefont {Jamet},
  \citenamefont {Jacquet}, \citenamefont {George}, \citenamefont
  {Barth{\'{e}}l{\'{e}}my}, \citenamefont {Jaffr{\`{e}}s}, \citenamefont
  {Fert}, \citenamefont {Bibes},\ and\ \citenamefont {Vila}}]{Lesne2016}%
  \BibitemOpen
  \bibfield  {author} {\bibinfo {author} {\bibfnamefont {E.}~\bibnamefont
  {Lesne}}, \bibinfo {author} {\bibfnamefont {Yu}~\bibnamefont {Fu}}, \bibinfo
  {author} {\bibfnamefont {S.}~\bibnamefont {Oyarzun}}, \bibinfo {author}
  {\bibfnamefont {J.~C.}\ \bibnamefont {Rojas-S{\'{a}}nchez}}, \bibinfo
  {author} {\bibfnamefont {D.~C.}\ \bibnamefont {Vaz}}, \bibinfo {author}
  {\bibfnamefont {H.}~\bibnamefont {Naganuma}}, \bibinfo {author}
  {\bibfnamefont {G.}~\bibnamefont {Sicoli}}, \bibinfo {author} {\bibfnamefont
  {J.~P.}\ \bibnamefont {Attan{\'{e}}}}, \bibinfo {author} {\bibfnamefont
  {M.}~\bibnamefont {Jamet}}, \bibinfo {author} {\bibfnamefont
  {E.}~\bibnamefont {Jacquet}}, \bibinfo {author} {\bibfnamefont {J.~M.}\
  \bibnamefont {George}}, \bibinfo {author} {\bibfnamefont {A.}~\bibnamefont
  {Barth{\'{e}}l{\'{e}}my}}, \bibinfo {author} {\bibfnamefont {H.}~\bibnamefont
  {Jaffr{\`{e}}s}}, \bibinfo {author} {\bibfnamefont {A.}~\bibnamefont {Fert}},
  \bibinfo {author} {\bibfnamefont {M.}~\bibnamefont {Bibes}}, \ and\ \bibinfo
  {author} {\bibfnamefont {L.}~\bibnamefont {Vila}},\ }\bibfield  {title}
  {\enquote {\bibinfo {title} {{Highly efficient and tunable spin-to-charge
  conversion through Rashba coupling at oxide interfaces}},}\ }\href {\doibase
  10.1038/nmat4726} {\bibfield  {journal} {\bibinfo  {journal} {Nature
  Materials}\ }\textbf {\bibinfo {volume} {15}},\ \bibinfo {pages} {1261}
  (\bibinfo {year} {2016})}\BibitemShut {NoStop}%
\bibitem [{\citenamefont {No{\"{e}}l}\ \emph {et~al.}(2020)\citenamefont
  {No{\"{e}}l}, \citenamefont {Trier}, \citenamefont {Arche}, \citenamefont
  {Br{\'{e}}hin}, \citenamefont {Vaz}, \citenamefont {Garcia}, \citenamefont
  {Fusil}, \citenamefont {Barth{\'{e}}l{\'{e}}my}, \citenamefont {Vila},
  \citenamefont {Bibes},\ and\ \citenamefont {Attan{\'{e}}}}]{Noel2020}%
  \BibitemOpen
  \bibfield  {author} {\bibinfo {author} {\bibfnamefont {Paul}\ \bibnamefont
  {No{\"{e}}l}}, \bibinfo {author} {\bibfnamefont {Felix}\ \bibnamefont
  {Trier}}, \bibinfo {author} {\bibfnamefont {Luis M~Vicente}\ \bibnamefont
  {Arche}}, \bibinfo {author} {\bibfnamefont {Julien}\ \bibnamefont
  {Br{\'{e}}hin}}, \bibinfo {author} {\bibfnamefont {Diogo~C}\ \bibnamefont
  {Vaz}}, \bibinfo {author} {\bibfnamefont {Vincent}\ \bibnamefont {Garcia}},
  \bibinfo {author} {\bibfnamefont {St{\'{e}}phane}\ \bibnamefont {Fusil}},
  \bibinfo {author} {\bibfnamefont {Agn{\`{e}}s}\ \bibnamefont
  {Barth{\'{e}}l{\'{e}}my}}, \bibinfo {author} {\bibfnamefont {Laurent}\
  \bibnamefont {Vila}}, \bibinfo {author} {\bibfnamefont {Manuel}\ \bibnamefont
  {Bibes}}, \ and\ \bibinfo {author} {\bibfnamefont {Jean-philippe}\
  \bibnamefont {Attan{\'{e}}}},\ }\bibfield  {title} {\enquote {\bibinfo
  {title} {{Non-volatile electric control of spin – charge conversion in a
  SrTiO 3 Rashba system}},}\ }\href {\doibase 10.1038/s41586-020-2197-9}
  {\bibfield  {journal} {\bibinfo  {journal} {Nature}\ }\textbf {\bibinfo
  {volume} {580}},\ \bibinfo {pages} {483} (\bibinfo {year}
  {2020})}\BibitemShut {NoStop}%
\bibitem [{\citenamefont {Afzal}\ \emph {et~al.}(2019)\citenamefont {Afzal},
  \citenamefont {Min}, \citenamefont {Ko},\ and\ \citenamefont
  {Eom}}]{Afzal2019}%
  \BibitemOpen
  \bibfield  {author} {\bibinfo {author} {\bibfnamefont {Amir~Muhammad}\
  \bibnamefont {Afzal}}, \bibinfo {author} {\bibfnamefont {Kuen~Hong}\
  \bibnamefont {Min}}, \bibinfo {author} {\bibfnamefont {Byung~Min}\
  \bibnamefont {Ko}}, \ and\ \bibinfo {author} {\bibfnamefont {Jonghwa}\
  \bibnamefont {Eom}},\ }\bibfield  {title} {\enquote {\bibinfo {title}
  {Observation of giant spin–orbit interaction in graphene and heavy metal
  heterostructures},}\ }\href {\doibase 10.1039/C9RA06961E} {\bibfield
  {journal} {\bibinfo  {journal} {RSC Adv.}\ }\textbf {\bibinfo {volume} {9}},\
  \bibinfo {pages} {31797--31805} (\bibinfo {year} {2019})}\BibitemShut
  {NoStop}%
\bibitem [{\citenamefont {Groth}\ \emph {et~al.}(2014)\citenamefont {Groth},
  \citenamefont {Wimmer}, \citenamefont {Akhmerov},\ and\ \citenamefont
  {Waintal}}]{Groth2014}%
  \BibitemOpen
  \bibfield  {author} {\bibinfo {author} {\bibfnamefont {Christoph~W.}\
  \bibnamefont {Groth}}, \bibinfo {author} {\bibfnamefont {Michael}\
  \bibnamefont {Wimmer}}, \bibinfo {author} {\bibfnamefont {Anton~R.}\
  \bibnamefont {Akhmerov}}, \ and\ \bibinfo {author} {\bibfnamefont {Xavier}\
  \bibnamefont {Waintal}},\ }\bibfield  {title} {\enquote {\bibinfo {title}
  {{Kwant: A software package for quantum transport}},}\ }\href {\doibase
  10.1088/1367-2630/16/6/063065} {\bibfield  {journal} {\bibinfo  {journal}
  {New Journal of Physics}\ }\textbf {\bibinfo {volume} {16}},\ \bibinfo
  {pages} {063065} (\bibinfo {year} {2014})},\ \Eprint
  {http://arxiv.org/abs/1309.2926} {arXiv:1309.2926} \BibitemShut {NoStop}%
\end{thebibliography}%

\end{document}